\documentclass[runningheads]{llncs}

\usepackage[T1]{fontenc}

\usepackage{pifont}
\usepackage{import}

\usepackage{amsmath}
\usepackage{amsfonts}

\usepackage[operators, sets, lambda]{cryptocode}

\usepackage{graphicx}

\usepackage{tikz}
\usepackage[most]{tcolorbox}

\usepackage[table]{xcolor}
\usepackage{transparent}

\usepackage{tabularx}
\usepackage{booktabs}

\usepackage{xurl}
\usepackage{hyperref}
\usepackage[htt]{hyphenat}
\usepackage{csquotes}
\usepackage[shortlabels, inline]{enumitem}
\usepackage[normalem]{ulem}
\usepackage{zebra-goodies}

\usepackage[backend=biber,style=lncs]{biblatex} 

\usepackage[capitalise, noabbrev]{cleveref}

\hypersetup{
linktoc=all,
  breaklinks=true,
  colorlinks=true,
  linkcolor=black,
  citecolor=black,
  urlcolor=blue
}

\renewcommand{\mkbegdispquote}[2]{\openautoquote}

\DeclareSourcemap{
  \maps[datatype=bibtex,overwrite]{
    \map{
      \step[fieldsource=doi,
        match={\\_},
      replace={_}]
      \step[fieldsource=url,
        match={\\_},
      replace={_}]
    }
  }
}

\definecolor{modernblue}{RGB}{59, 130, 246}
\definecolor{moderngray}{RGB}{107, 114, 128}
\definecolor{lightgray}{RGB}{237, 239, 241}
\definecolor{darkgray}{RGB}{43, 54, 71}
\definecolor{bordercolor}{RGB}{217, 220, 225}
\tcbset{
    colback=lightgray,
colframe=bordercolor,
    arc=0pt,
outer arc=0pt,
boxrule=0.5pt,
    left=8pt,
    right=8pt,
    top=6pt,
    bottom=6pt,
coltitle=black!90,
enhanced,
    breakable,
}
\newtcbtheorem[crefname={Protocol}{Protocol},Crefname={Protocol}{Protocol}]{protocol}{Protocol}{}{proto}
\crefformat{tcb@cnt@Protocol}{Protocol~#2#1#3}
\Crefformat{tcb@cnt@Protocol}{Protocol~#2#1#3}

\newenvironment{enum}{
\begin{enumerate*}[{\bfseries (\arabic*)}]}{
\end{enumerate*}}

\newcommand{\BigSetInAlign}[2]{&\left\{
        \begin{aligned}#1
        \end{aligned} \right. \\
        &\quad \left. \vphantom{
            \begin{aligned}#1
        \end{aligned}}
        \middle|\ \,
        \begin{aligned}#2
    \end{aligned}\right\}
}

\newcommand{\rem}{\,\%\,}
\newcommand{\mlex}{\mathsf{mlex}}
\newcommand{\unex}{\mathsf{unex}}
\newcommand{\mlin}{\mathsf{mlin}}
\newcommand{\eq}{K}

\newcommand{\instance}{\mathbf{x}}
\newcommand{\witness}{\mathbf{w}}
\newcommand{\oracle}[1]{\mathcal{O}^{#1}}
\newcommand{\sumPoly}{P}
\newcommand{\roundPoly}{p}
\newcommand{\sq}{\mathrm{sq}}
\newcommand{\no}{\mathrm{no}}
\newcommand{\even}{\mathrm{ev}}
\newcommand{\odd}{\mathrm{od}}
\newcommand{\degBound}{D}

\begin{document}

\title{Protocols for Univariate Sumcheck}

\author{Malcom Mohamed}\institute{\email{m@mal.com.de}\\
\url{https://mal.com.de}}

\date{\today}

\maketitle

\begin{abstract}
  Two candidate approaches for univariate sumcheck over roots of
  unity are presented. The first takes the form of a multilinear
  evaluation protocol, which can be combined with the standard
  multivariate sumcheck protocol. The other consists of a direct
  reduction from univariate sumcheck to multilinear evaluation, which
  can be combined with Gemini (Bootle et al., Eurocrypt 2022).
  Optionally, natural round reductions from $m$ to $\log(m)$ or
  $O(\sqrt{m})$ are supported, while retaining linear prover time.
\end{abstract}

\section{Introduction}\label{sec:intro}
SNARKs commonly encode data as polynomials and then use algebraic techniques to prove facts about the data in the sense of being in an NP language.
\emph{Sumcheck} has emerged as a core building block of such systems \cite{sumcheckArguments}.
Sumcheck here refers to any probabilistic reduction of the type \enquote{$\sum_i \sumPoly(i) = s$ if and only if $\sumPoly'(r)=t$} where $\sumPoly'$ is related to $\sumPoly$ and $\sumPoly'$, $r$ and $t$ are determined by the reduction.

We let $v$ be a vector and, given the SNARK setting, may interpret it as the satisfying variable assignment for a given system of constraint equations.
We denote its multilinear extension polynomial as $\mlex[v]$ and its univariate extension polynomial as $\unex[v]$.
The two extensions give rise to a separation of the SNARK landscape into univariate systems that internally use $\unex$ to encode their data and multilinear ones that use $\mlex$.
The starkest cryptographic design difference between these classes of systems comes from the kinds of sumcheck protocols that can be used with each.
The standard multivariate sumcheck \cite{lfkn} is typically favored for prover efficiency while the standard univariate protocol \cite{aurora} takes just one round and can thus be considered as more bandwidth and verifier friendly.
It is a natural question whether there is a univariate sumcheck protocol that matches the prover efficiency of the multivariate one.

Motivated by this apparent separation, researchers have proposed adaptors which identify multilinear polynomials with univariate polynomials in order to bring \enquote{opposite world} techniques into other systems.
These adaptors can be categorized by\begin{enum}
\item how \enquote{opposite world} polynomials are mapped to the \enquote{own world} and
\item how statements about the mapped polynomials are proven.\end{enum}
The known categories along dimension \textbf{(1)} are mapping coefficients to coefficients~\cite{gemini}, values to coefficients~\cite{hyperkzg,zeromorph} and values to values~\cite{dualpcs,logupgkr}. As methods for \textbf{(2)}, we are aware of recursive folding~\cite{gemini,hyperkzg}, quotients~\cite{zeromorph,dualpcs} and kernel interpolation approaches~\cite{logupgkr}.
See \cref{tab:categories} for the resulting classification and representative references in the multivariate-to-univariate direction.\footnote{There are fewer cases of the other direction so far, though we can point to~\cite[Section 5]{lookupAdaptor} and~\cite{biniusNote} for values-to-values kernel interpolation approaches.}
We stress that the values-to-values mapping is especially interesting because it brings with it all standard \emph{sumcheck} techniques from the \enquote{opposite world} without any overhead for changing between coefficient and evaluation basis.
See~\cite[Section 5]{hybridplonk} for an example of the non-trivial extra steps required for such basis translation.

\begin{table}[!t]
    \centering
    \caption{Maps from multilinears to univariates that enable \enquote{opposite world} sumchecks.}\label{tab:categories}
    {\footnotesize\rowcolors{2}{white}{lightgray}
\begin{tabularx}{\textwidth}{p{2cm}Xp{3.4cm}p{2.6cm}}
            \toprule
            & Coefficients to coefficients & Values to coefficients & Values to values \\
            \midrule
            Recursive\newline{}folding & Gemini tensor protocol\newline\cite[Sec. 2.4.2]{gemini} \cite[Sec. 5.1]{geminiEprint} & HyperKZG \cite[Sec. 6]{hyperkzg} & \textbf{This article}\newline{}(\cref{proto:mlex2unex}) \\
            Quotients & Caulk \cite[Sec. 4.6]{caulk} & Zeromorph \cite{zeromorph} & Ganesh et al. \cite{dualpcs} \\
            Kernel\newline{}interpolation & & & Papini-Haböck\newline\cite[Sec. 5.1]{logupgkr} \\
            \bottomrule
    \end{tabularx}}
\end{table}
 
Besides using a multivariate-to-univariate adaptor in order to apply the multivariate sumcheck protocol to univariates, there is also the question of whether univariate polynomials have direct linear-time sumchecks.
It is a folklore belief that the Gemini (adaptor) protocol itself can be used as such:\footnote{Also \cite[Section 1.2]{timespaceSumcheck}: ``Gemini [...] proposes a [...] sumcheck protocol for univariate polynomials''.}
\begingroup
\setlength{\leftmargini}{2em}
\setlength{\rightmargin}{2em}
\begin{displayquote}[{\cite[Remark 5]{gemini} \cite[Remark 2.8]{geminiEprint}}]
    Based on [the Gemini adaptor], one can construct a linear-time univariate sumcheck protocol [...], which we believe could be of independent interest for future research.
    There are other univariate sumcheck protocols in the literature; however, these protocols cannot be used in our setting.
    [...]
    [A protocol of Drake's] may conceivably lead to a univariate sumcheck protocol [...]
    However, no details (or implementations) of the protocol are available.
\end{displayquote}
\endgroup
But as far as we know,
no linear-time univariate sumcheck protocol based on Gemini has since been made public. The protocol ascribed to Drake in the quoted remark has since only been documented in a blog post that circulated in the community~\cite{dgm} (henceforth DGM after its authors Drake, Gabizon and Meckler).

In this context, the present article makes the following contributions.

\paragraph{Reducing $\mlex$ to $\unex$.}
\cref{sec:mlex2unex} shows a linear-time recursive folding protocol for $\mlex$-to-$\unex$ identification.
The protocol allows using the standard multivariate sumcheck protocol for a given univariate extension polynomial in a straightforward manner.
By comparison, prior candidates like the one from~\cite[Section 5.1]{logupgkr} and the one implicit in~\cite{dualpcs} have at least quasilinear prover time.

\paragraph{Univariate sumcheck without $\mlex$.}
As an alternative to the multivariate sumcheck protocol, \cref{sec:dgm} describes a direct linear-time sumcheck for univariates. First, we clarify that the DGM protocol contains a flaw that invalidates it as written.
After that, we show that an adaptation of a fixed version of DGM is, in fact, an instance of the elusive Gemini-based linear-time univariate sumcheck. 

\paragraph{Round reductions.}
Both the composition of the multivariate sumcheck protocol with the new adaptor and the DGM-based univariate reduction admit natural round reductions (\cref{sec:round_reduction,sec:dgm_sumcheck}).
This is because in each case, stopping early after $k$ rounds leaves a smaller \emph{univariate} claim.
In the style of recent works~\cite{hybridplonk,timespaceSumcheck}, the small claim can be handled with a single-round sumcheck like Aurora's~\cite{aurora}, allowing for $\log(m)$-round executions.
Our protocols have \emph{no} overhead for this transformation, unlike~\cite{hybridplonk}, and do not depend on special non-standard interpolation domains, unlike~\cite[Section 5]{gruen} or~\cite[Section 7]{timespaceSumcheck}.

\newcommand{\myHspace}{\hphantom{-}}
\begin{table}[!t]
    \centering
    \caption{Protocols for proving $\sum_{i \in [0,2^m-1]} g(f_1(w^i),\allowbreak{} \dots,\allowbreak{} f_q(w^i)) = s$, where $g$ has degree $d$ and $\binom{d+q}{q}$ is a general bound for the time to evaluate $g$.
        LFKN stands for the standard multivariate sumcheck protocol~\cite{lfkn}.
        Columns 2 and 3 show time complexities in asymptotic notation (not counting the verifier's oracle queries).
        Columns 5 and 6 show the number of sent field elements and oracles.
\label{tab:summary}}
    {\footnotesize\rowcolors{2}{white}{lightgray}
        \begin{tabularx}{\textwidth}{p{1.9cm}Xp{1.9cm}p{1.7cm}p{2.15cm}p{1.2cm}}
            \toprule
            & Prover & Verifier & Rounds & $\FF$ & Oracles \\
            \midrule
            Aurora \cite{aurora} & $\left(q m + \binom{d+q}{q}\right) d 2^m$ & $\binom{d+q}{q}+m$ & $1$ & $0$ & $2$ \\
\midrule
            LFKN\newline{}+ \cref{proto:mlex2unex} & $\binom{d+q}{q} d 2^m$ & $dm+\binom{d+q}{q}$ & $m+1$ & $(d+1)m+q$ & $2m$ \\
            LFKN\newline{}+ \cref{proto:mlex2unex}\newline{}+ Aurora & $\binom{d+q}{q} d 2^m$ & $d\log(m)$ \newline{} $+ \binom{d+q}{q}+ m$ & $\log(m)+1$ & $(d+1)\log(m)$ & $2\log(m)$\newline{}$+q+2$ \\
\midrule
            \cref{proto:dgm}\newline{}+ Gemini & $\binom{d+2q}{2q} d 2^m$ & $d^2 m + \binom{d+q}{q}$ & $m+1$ & $O(d^2)m+q$ & $4m-1$ \\
            \cref{proto:dgm}\newline{}+ Gemini\newline{}+ Aurora & $\binom{d+2q}{2q} d 2^m$ & $d^2\log(m)$\newline{} $+ \binom{d+q}{q} + m$ & $\log(m)+1$ & $O(d^2) \log(m)$ & $4\log(m)$\newline{}$+q+1$ \\
\bottomrule
    \end{tabularx}}
\end{table}
 
\paragraph{Implications.}
\cref{tab:summary} contrasts the newly enabled protocols with the standard Aurora approach.
We note that in rows 2, 4 and 6, other adaptors could be plugged in and might yield different performance characteristics.
However, this might not be possible for row 3 and 5's round-reduced variants or might require bespoke adaptations.

The upshot is that linear-time univariate sumcheck---for pre-existing $\unex$ polynomials---\emph{is} generically achievable by combining a suitable reduction and an adaptor.
Thus, any current use case of the Aurora protocol may benefit from the prover efficiency previously reserved for the world of multivariate data encodings.

 \section{Preliminaries}\label{sec:prelims}
\subsection{Notations} $[a]$ denotes the set of integers $\{1, \dots, a\}$.
$[a, b]$ denotes the set of integers $\{a, \dots, b\}$.
$\FF$ denotes a finite field.
$w$ denotes a primitive $2^m$th root of unity in $\FF$, so that $\left\{w^i \mid i \in [0,2^m-1]\right\}$ is a multiplicative subgroup.
For $v \in \FF^{2^k}$, $\unex[v]$ denotes the Lagrange interpolation polynomial over the domain $\{w^{2^{m-k} i} \mid i \in [0,2^k-1]\}$ (\emph{univariate extension}) and $\mlex[v]$ denotes the Lagrange interpolation polynomial over $\{0,1\}^k$ (\emph{multilinear extension}).
That is, $\unex[v](w^{2^{m-k} i}) = \mlex[v](i_1, \dots, i_k) = v_i$ for all $i \in [0,2^k-1]$ with bits $i_1, \dots, i_k$---the least significant bit being $i_1$.

If $p$ is a univariate polynomial in $x$, then $\deg(p)$ denotes the degree of $p$ and $p[i]$ denotes the coefficient before $x^i$.
$\mlin$ denotes the inverse Kronecker substitution that re-interprets $p \in \FF[x]$ of degree at most $2^m-1$ as the multilinear polynomial with the same coefficients, $\mlin[p] = \sum_{i_1, \dots, i_m \in \{0,1\}} p[i_1 + \dots + 2^{m-1} i_m]\allowbreak{} x_1^{i_1} \cdots x_m^{i_m}.$ 

$p \rem q$ denotes the remainder of $p$ modulo $q$.

\subsection{Polynomial Interactive Oracle Proofs and Reductions}
\emph{Polynomial interactive oracle proofs (PIOPs)} are two-party protocols between a prover and a verifier which are defined for a given NP relation $R$ and a mathematical field with a polynomial ring.
The prover's input is an instance-witness pair $(\instance;\witness)$ of the NP relation, while the verifier starts only with $\instance$ and must output a bit.
We think of the prover as making the \emph{claim} that there exists $\witness$ such that $(\instance;\witness) \in R$.
The messages sent between prover and verifier are field elements or so-called polynomial oracles, denoted $\oracle{(\cdot)}$, idealized objects which provide black-box query access to polynomial functions.
We implicitly assume that there is a fixed upper bound $\degBound$ on the degree of oracle polynomials.
The core properties of a PIOP are completeness and soundness.
By \emph{(perfect) completeness}, a prover with a valid instance-witness pair in its input causes the verifier to output 1.
By \emph{soundness}, for any probabilistic polynomial-time adversarial prover, for any $\instance$ for which there is no $\witness$ such that $(\instance;\witness)\in R$, the verifier with input $\instance$ outputs 0 except with negligible probability (\emph{soundness error}).
For details on PIOPs, also refer to~\cite[Section 2.1]{hyperplonk}.

Related to the notion of a PIOP is that of a \emph{reduction}, by which we mean a protocol where the verifier outputs an instance of another NP relation and the prover outputs an instance-witness pair.
Here, completeness means that a valid first instance-witness pair leads to a valid second instance-witness pair.
Soundness means that if the first instance has no valid witness, neither will the second.
For a formalization of reductions, also refer to \cite{reductions,arc}.
Note that in this article, we allow polynomial oracles in reductions, too, even though it might be slightly non-standard.

\paragraph{The multivariate sumcheck protocol.}
While the multivariate sumcheck protocol dates back to \cite{lfkn}, our framing as a (reduction-based) PIOP is influenced by recent works like \cite{hyperplonk,reductions}.
The protocol uses the following recursive reduction from the relation
\[
    R = \left\{\left(\oracle{\sumPoly}, s; \sumPoly\right) \;\middle|\; \sum_{i_1 \in H_1, \dots, i_m \in H_m} \sumPoly(i_1,\dots,i_m) = s \right\}
\]
to (after $m$ rounds)
\[
    R' = \left\{\left(\oracle{\sumPoly}, r_1, \dots, r_m, s'; \sumPoly\right) \mid \sumPoly(r_1, \dots, r_m) = s'\right\}.
\]
Here, $\sumPoly$ is an $m$-variate polynomial, $s$ is a field element and $H_1, \dots, H_m$ are sets:

\begin{protocol}{Multivariate sumcheck (reduction step)}{sumcheck}

    \textbf{Claim:} $\sum_{i_1 \in H_1, \dots, i_m \in H_m} \sumPoly(i_1, \dots, i_m) = s$.
    \begin{enumerate}
        \item The prover sends
            \[
                \roundPoly_1(y) = \sum_{i_2 \in H_2, \dots, i_m \in H_m} \sumPoly(y, i_2, \dots, i_m).
            \]
\item The verifier checks that $\sum_{i_1 \in H_1} \roundPoly_1(i_1) = s$.
        \item The verifier sends $r_1$, chosen at random.
        \item Both prover and verifier set $s' = \roundPoly_1(r_1)$.
    \end{enumerate}
    \textbf{New claim:} $\sum_{i_2 \in H_2, \dots, i_m \in H_m} \sumPoly(r_1, i_2, \dots, i_m) = s'$.
\end{protocol}

The idea of the sumcheck protocol is to handle the \emph{new claim} with a recursive invocation of the reduction.
After $m$ rounds, the claim will have been reduced to $\sumPoly(r_1,\allowbreak{} \dots,\allowbreak{} r_m) = s'$.
This claim is then checked with a single query to $P$.

\paragraph{Gemini.}
In this article, Gemini (to be exact, only its ``tensor product'' subprotocol) is understood as a coefficient-to-coefficient multilinear-to-univariate adaptor, that is, a protocol to prove statements of the form ``$\mlin[f](z_1,\allowbreak{} \dots,\allowbreak{} z_m) = s$'' to a verifier with oracle access to the univariate polynomial $f$ \cite[cf.][]{gemini}.
Internally, Gemini is based on the even-odd decomposition of $f$: the fact that there are unique polynomials $f_{\even}$ and $f_{\odd}$ of degree at most $\floor{\frac{\deg(f)}{2}}$---consisting of the even and, respectively, odd coefficients of $f$---such that $f(x) = f_{\even}(x^2) + x f_{\odd}(x^2)$, $f_{\even}(x^2) = \frac{f(x)+f(-x)}{2}$ and $f_{\odd}(x^2) = \frac{f(x)-f(-x)}{2x}.$
Gemini combines these properties with the observation that $\mlin[f](x_1,\allowbreak{} \dots,\allowbreak{} x_m) = \mlin[f_{\even}](x_2,\allowbreak{} \dots,\allowbreak{} x_m) + x_1 \mlin[f_{\odd}](x_2,\allowbreak{} \dots,\allowbreak{} x_m).$
The protocol works by the prover sending univariate polynomials $f_1,\allowbreak{}\dots,\allowbreak{}f_{m-1}$ where $\mlin[f_i] = \mlin[f](z_1,\allowbreak{}\dots,\allowbreak{}z_i,\allowbreak{}x_{i+1},\allowbreak{}\dots,\allowbreak{}x_m)$ and the verifier checking each one based on the previous, using the above identities.
The prover runs in time in $O(2^m)$ and sends $m-1$ oracles.
After that, the verifier makes $3m-1$ total queries to the oracles (and to the original $f$ oracle) at points $\pm r$ and $r^2$ for a random $r$.
 \section{Evaluating $\mlex$ via $\unex$}\label{sec:mlex2unex}

This section describes a univariate PIOP for proving the statement \enquote{$\mlex[v](z_1,\allowbreak{}\dots,\allowbreak{}z_m)=s$} to a verifier who has access to a $\unex[v]$ oracle.

\subsection{Square Evaluation Folding}

Our protocol starts with the prover providing an oracle for $\unex[v']$ and claiming that $v'$ has half the length of $v$ and that $\mlex[v'](x_2,\allowbreak{}\dots,\allowbreak{}x_m) = \mlex[v](z_1,\allowbreak{} x_2,\allowbreak{} \dots,\allowbreak{} x_m)$.
Ignoring for a moment how that claim is checked, the protocol will then recurse on the reduced claim \enquote{$\mlex[v'](z_2,\allowbreak{}\dots,\allowbreak{}z_m)=s$}.
After $m$ recursive rounds, the final $\unex$ polynomial is the constant $\mlex[v](z_1,\allowbreak{}\dots,\allowbreak{}z_m)$.
The verifier directly tests equality with $s$ and we are done.

To let the verifier confirm that $\unex[v']$ really corresponds to the partial $\mlex$ evaluation, we make use of $\{1,\allowbreak{} w^2,\allowbreak{} \dots,\allowbreak{} w^{2^m-2}\}$ as the interpolation domain of $\unex[v']$ and will now derive a helpful polynomial decomposition.
Denote $\unex[v]=f$ and $\unex[v'] = f'$, so that $v = (f(1),\allowbreak{} f(w),\allowbreak{} \dots,\allowbreak{} f(w^{2^m-1}))$ and $v'=(f'(1),\allowbreak{} f'(w^2),\allowbreak{} \dots,\allowbreak{} f'(w^{2^m-2}))$.
By interpolation, we know that
\begin{align*}
    \mlex[v'](x_2,\dots,x_m)
&= (1-z_1) \mlex[v](0, x_2, \dots, x_m) \\
    &\quad + z_1 \mlex[v](1,x_2,\dots,x_m).
\end{align*}
Thus, for all $i \in [0,2^{m-1}-1]$, we must enforce that \begin{align*}
    f'(w^{2i})
&= (1-z_1) f(w^{2i}) + z_1 f(w^{2i+1}).
\end{align*}
This requirement motivates the following decomposition of $f$ into a \emph{square part} $f_{\sq}$ and a \emph{non-square part} $f_\no$ such that $\deg(f_\sq),\allowbreak{} \deg(f_\no) \leq 2^{m-1}-1$ and
\begin{align*}
    f_\sq(w^{2i}) &= f(w^{2i}) \\
    f_\no(w^{2i}) &= f(w^{2i+1})
\end{align*}
for all $i \in [0,2^{m-1}-1]$.
This is equivalent to saying that
\begin{align*}
    f_\sq(x) &= f \rem (x^{2^{m-1}} - 1) \\
    f_\no(x) &= \left(f \rem (x^{2^{m-1}} + 1)\right)(wx)
\end{align*}
since the remainder of any $p$ modulo $q$, where $q$ is monic, is the unique polynomial of degree below $\deg(q)$ that agrees with $p$ on the zeros of $q$.

With the square decomposition, we can rephrase the verifier's goal as confirming that $f' = (1-z_1) f_\sq + z_1 f_\no$.
This, in turn, is straightforwardly possible with the following generalized Lagrange interpolation formula:
\begin{equation}\label{eq:crt_interpolation}
    f(x) = \frac{1+x^{2^{m-1}}}{2} f_\sq(x) + \frac{1-x^{2^{m-1}}}{2} f_\no(w^{-1}x).
\end{equation}
The above equality holds because both sides define degree-$(2^m-1)$ polynomials that agree on all $2^m$ roots of unity.

To make use of \cref{eq:crt_interpolation}, our protocol (\cref{proto:mlex2unex}) has the prover send $f_\sq$ and $f_\no$.
The verifier checks them against $f$ by testing \cref{eq:crt_interpolation} at a random point.
Convinced that $f_\sq$ and $f_\no$ are the unique square and non-square components of $f$, the verifier can from now on query them to compute the correct value of $f'$ at any point.
For $j \in [0,\allowbreak{} m-1]$, denote as $f_j$ the $j$th folded polynomial, that is, the univariate extension over the set of all $w^{2^j i}$ of the length-$2^{m-j}$ evaluation vector of $\mlex[v](z_1,\allowbreak{} \dots,\allowbreak{} z_j,\allowbreak{} x_{j+1},\allowbreak{} \dots,\allowbreak{} x_m)$.

\begin{protocol}{$\mlex$-to-$\unex$}{mlex2unex}
    \textbf{Claim:} $\mlex[v](z_1,\dots,z_m)=s$.
    \begin{enumerate}
        \item The prover sends oracles for $f_{j,\sq}$ and $f_{j,\no}$ for $j \in [0,m-1]$.
        \item The verifier samples $r$ at random and queries $f_0(r)$ and $f_{j,\sq}(r),\allowbreak{} f_{j,\no}(r),\allowbreak{} f_{j,\no}(w^{-2^j}r)$ for $j \in [0,m-1]$.
        \item The verifier checks
\[
                f_0(r) = \frac{1+r^{2^{m-1}}}{2} f_{0,\sq}(r) + \frac{1-r^{2^{m-1}}}{2} f_{0,\no}(w^{-1}r)
            \]
            and, for all $j \in [m-1]$,
            \begin{align*}
                &(1-z_j) f_{j-1,\sq}(r) + z_j f_{j-1,\no}(r) \\
                =& \frac{1+r^{2^{m-j-1}}}{2} f_{j,\sq}(r)
                + \frac{1-r^{2^{m-j-1}}}{2} f_{j,\no}(w^{-2^j}r)
            \end{align*}
            and
            \[
                (1-z_m) f_{m-1,\sq}(r) + z_m f_{m-1,\no}(r) = s.
            \]
    \end{enumerate}
\end{protocol}

\begin{proposition}\label{prop:mlex2unex}
    \cref{proto:mlex2unex} is a PIOP for the relation
    \[
        R = \left\{
            \left(\oracle{f_0}, z_1, \dots, z_m, s; v\right)
            \,\middle|\,
            f_0 = \unex[v] \land \mlex[v](z_1, \dots, z_m) = s
        \right\}
    \]
    with
    \begin{itemize}
        \item perfect completeness,
        \item soundness error in $O(\frac{\degBound}{|\FF|})$ (where $\degBound$ is a global upper bound on any oracle polynomial's degree), \item prover running time in $O(2^m)$,
        \item verifier running time in $O(m)$,
        \item 1 round,
        \item no field elements sent by the prover,
        \item $2m$ oracles sent by the prover,
        \item $3m+1$ oracle queries made by the verifier ($2m+1$ of which at the same point).
    \end{itemize}
\end{proposition}

\begin{proof}
    The properties besides completeness, soundness and prover run time are obvious from the description of \cref{proto:mlex2unex}.

    \paragraph{Completeness.}
    Suppose that $\mlex[v](z_1,\allowbreak{} \dots,\allowbreak{} z_m) = s$ and the prover is honest.
    Since the prover is honest, each $f_{j,\sq}$ and $f_{j,\no}$ are indeed the square and, respectively, non-square components of $f_j$. Then \cref{eq:crt_interpolation} ensures completeness: first, it directly implies that the verifier's first check must hold.
    For each $j$th subsequent check, observe that the left-hand side is exactly $f_j(r)$.
    Therefore, the round-$j$ version of \cref{eq:crt_interpolation} for the $2^{m-j}$th roots of unity ensures that the check holds.

    \paragraph{Soundness.}
    Consider an execution of \cref{proto:mlex2unex} where the verifier accepts although $\mlex[v](z_1,\allowbreak{} \dots,\allowbreak{} z_m) \neq s$.
    Then there must exist a round $k \in [0,m-1]$ where at least one of the prover's sent oracles $f_{k,\sq},\allowbreak{} f_{k,\no}$ is \emph{not} the square or, respectively, non-square part of $f_{k} = (1-z_k) f_{k-1,\sq} + z_k f_{k-1,\no}$.
    The verifier's corresponding check,
    \[
        f_{k}(r) = \frac{1+r^{2^{m-k-1}}}{2} f_{k,\sq}(r) + \frac{1-r^{2^{m-k-1}}}{2} f_{k,\no}(w^{-2^{k}}r),
    \]
    tests \cref{eq:crt_interpolation} at the independently sampled point $r$.
    By Schwartz-Zippel, this check passes with probability at most $\frac{\degBound}{|\FF|}$ over the choice of $r$ (where $\degBound$ bounds every oracle's degree).
    Therefore, the malicious prover's overall success probability is also at most $\frac{\degBound}{|\FF|}$.

    \paragraph{Prover time.}
    In order for the prover to send the required oracles, the prover must internally compute explicit representations of the polynomials.
    Since $f_{j,\sq}$ and $f_{j,\no}$ have degree at most $2^{m-j-1}-1$, an explicit representation of each is given by the $2^{m-j-1}$ evaluations on the set of all $w^{2^{j+1} i}$.
    From the values of $f_{j-1}$, which are also the required values of $f_{j-1,\sq}$ and $f_{j-1,\no}$, the values of $f_j = (1-z_j) f_{j-1,\sq} + z_j f_{j-1,\no}$ can be computed in one pass with $O(2^{m-j})$ operations.
    Summing over all $j$, this makes for a total running time in $O(2^m)$.

    \paragraph{}
    This completes the proof of \cref{prop:mlex2unex}.
\end{proof}

\subsection{Comparisons}

Papini and Haböck \cite[Section 5]{logupgkr} propose a kernel interpolation approach for $\mlex$-to-$\unex$ identification.
That is, they write the evaluation $\mlex[v](z_1,\allowbreak{}, \dots,\allowbreak{} z_m)$ as the sum $\sum_i \unex[v](w^i) \eq(i_1,\allowbreak{} \dots,\allowbreak{} i_m,\allowbreak{} z_1,\allowbreak{} \dots,\allowbreak{} z_m)$ where $\eq$ is the multilinear interpolation kernel. Their protocol proceeds by sending a $\unex$ oracle to the interpolant of the values of $\eq$.
Then, a special sumcheck is formulated in order to check that the oracle indeed contains $\eq$ \cite[Section A.1]{logupgkr}.
This sumcheck claim and the main interpolation claim are both handled by the Aurora sumcheck protocol \cite{aurora}, costing $O(m2^m)$ proving time. Despite sending fewer oracles, the protocol still needs $m+1$ queries to the $\unex$ corresponding to $\eq$.

Ganesh et al.'s work \cite[Section 4]{dualpcs} contains a quotienting solution for $\mlex$-$\unex$ identification.
We may briefly re-interpret their description as a KZG-based commitment scheme as the result of compiling an underlying PIOP. Ganesh et al. define the \emph{linear map} that takes $\mlex[v]$ to $\unex[v]$ as $U$ and apply $U$ to the fact that $\mlex[v](z_1,\allowbreak{}, \dots,\allowbreak{} z_m)=s$ if and only if there exist $q_1,\allowbreak{} \dots,\allowbreak{} q_m$ such that $\mlex[v] = s + \sum_j q_j \cdot (x_j-z_j)$.
Their protocol must overcome the fact that $U$ lacks a simple algebraic description.
Prover run time is in $O(m^2 2^m)$ due to $O(m)$ required FFTs. The protocol takes 10 rounds of interaction and sends $2m$ intermediary oracles and $6m-3$ field elements from prover to verifier. These messages have the purpose of helping the verifier test the aforementioned identity at a random point under $U$. For a similar reason, the protocol also requires a trusted party to hand the verifier oracles to certain fixed polynomials upfront.

Gemini and our \cref{proto:mlex2unex} solve different problems but are structurally similar.
For what it does, Gemini is more communication-efficient than our adaptor since it only sends one oracle for every two sent by \cref{proto:mlex2unex} for each $j \in [0,m-1]$.
Gemini also performs 2 fewer oracle queries.

\subsection{Sumcheck and Round Reduction}\label{sec:round_reduction}

The main application of a $\mlex$-to-$\unex$ adaptor is running the multivariate sumcheck protocol against univariate oracles.
A number of recent works proposes \emph{round reductions} for multivariate sumcheck~\cite[Section 5]{gruen}~\cite{hybridplonk,levrat,logstar,timespaceSumcheck}.
It is easy to see that our adaptor makes a round reduction from $m$ to $\log(m)$ completely natural, meeting~\cite{hybridplonk,levrat} with minimal effort.

\paragraph{Normal adaptor usage.}
If a verifier has access to the univariate oracles $f_1, \dots, f_q$ where $f_i = \unex[v_i]$ that determine $P=g(f_1,\allowbreak{} \dots,\allowbreak{} f_q)$, an adaptor protocol like ours can be used to prove a univariate sumcheck claim \enquote{$\sum_i P(i)=s$} as follows.
First, prover and verifier use the multivariate protocol on $P'=g(\mlex[v_1],\allowbreak{} \dots,\allowbreak{} \mlex[v_q])$ to reduce the claim to evaluation claims about $\mlex[v_i]$.
Then, these claims---or preferably a single batched claim about a random linear combination---are handled with the adaptor.
The costs of this approach are shown in the second row of \cref{tab:summary}.

\paragraph{Round reduction.}
Perhaps unsurprisingly, reducing the number of rounds can be achieved by early-stopping the multivariate protocol after $k<m$ rounds.
Once the claim about $P$ has been reduced to a smaller sumcheck on $P(r_1,\allowbreak{} \dots,\allowbreak{} r_k,\allowbreak{} x_{k+1},\allowbreak{} \dots,\allowbreak{} x_m)$, a separate \emph{single-round} protocol like Aurora's~\cite{aurora} may be used.
We observe that \cref{proto:mlex2unex} is readily compatible with this kind of early stopping.
Indeed, our proof of \cref{prop:mlex2unex} implies that the verifier's first $k$ checks suffice to authenticate a newly provided $\unex$ oracle as corresponding to a partial evaluation $\mlex[v](z_1,\allowbreak{} \dots,\allowbreak{} z_k,\allowbreak{} x_{k+1},\allowbreak{} \dots,\allowbreak{} x_m)$.
Performing these $k$ steps of the adaptor for $f_1,\allowbreak{} \dots,\allowbreak{} f_q$ thus creates \emph{exactly} the starting conditions for a univariate sumcheck.
The round number can now be chosen, for example, as $k = \log(m)$ to guarantee that even a quasilinear single-round sumcheck is actually in $O(2^m)$.
This is because it will be quasilinear in $2^{m-k}$ instead of $2^m$ and
\begin{align*}
    (m-\log(m)) 2^{m-\log(m)}
    &= 2^m \cdot (m-\log(m)) 2^{-\log(m)} \\
    &= 2^m \cdot \frac{m-\log(m)}{m} \\
    &= 2^m \cdot \left(1-\frac{\log(m)}{m}\right)
    < 2^m.
\end{align*}
The third row of \cref{tab:summary} shows the costs of the resulting $\log(m)$-round protocol.

\begin{remark}
    The described approach can be seen as a remarkable simplification of HybridPlonk \cite{hybridplonk}, which also cuts the last $m-k$ sumcheck rounds short using a univariate sumcheck.
    However, HybridPlonk is based on a \emph{values-to-coefficients} mapping of multivariates to univariates.
    This necessitates an extra basis translation phase \cite[Section 5]{hybridplonk} that this article's method entirely eliminates.
\end{remark}

\begin{remark}
    It remains to be seen if our (or any) adaptor can also seamlessly reduce rounds beyond $\log(m)$ as in \cite{logstar,timespaceSumcheck} while preserving the convenient direct mapping from the Boolean hypercube to the roots-of-unity domain(s).
    As in \cite[Section 5]{gruen}, it might also be desirable to perform the \enquote{big} sumcheck at the start rather than at the end.
    This is because the domain values will not yet have been mixed with the protocol's randomness, possibly resulting in cheaper arithmetic operations when the original values are small.
\end{remark}

 \section{Reducing Sumcheck to
$\mlin$}\label{sec:dgm}\label{sec:gemini_reduction}

This section describes a univariate domain identity PIOP (and thereby
also a sumcheck) based on DGM's ideas to reduce to Gemini.

\subsection{DGM Recap}\label{sec:dgm_recap}

DGM \cite{dgm} is a proposed protocol for domain identity, that is,
proving that $\sumPoly(w^i) = h(w^i)$ for all $i \in [0,\allowbreak{}
2^m-1]$ where $\sumPoly = g(f_1,\allowbreak{} \dots,\allowbreak{}
f_q)$, $g$ is a degree-$d$ polynomial and where the verifier has
oracle access to the degree-$(2^m-1)$ polynomials
$f_1,\allowbreak{}\dots,\allowbreak{}f_q$ and $h$.
The protocol is based on the following two main ideas.

First, for any polynomials $p$ and $p'$, we have $p(w^i) = p'(w^i)$
for all $i \in [0,2^m-1]$ if and only if $p_{\even}(w^{2i}) =
p'_{\even}(w^{2i})$ and $p_{\odd}(w^{2i}) = p'_{\odd}(w^{2i})$ for
all $i \in [0,2^{m-1}-1]$.
To see this, consider that if $p$ and $p'$ agree on all $w^i$, then
\[
  p_{\even}(w^{2i})
  = \frac{p(w^i)+p(-w^i)}{2}
  = \frac{p'(w^i)+p'(-w^i)}{2}
  = p'_{\even}(w^{2i})
\]
and
\[
  p_{\odd}(w^{2i})
  = \frac{p(w^i)-p(-w^i)}{2w^i}
  = \frac{p'(w^i)-p'(-w^i)}{2w^i}
  = p'_{\odd}(w^{2i}).
\]
Conversely, if the even and odd parts agree on all $w^{2i}$, then
\[
  p(w^i)
  = p_{\even}(w^{2i}) + w^i p_{\odd}(w^{2i})
  = p'_{\even}(w^{2i}) + w^i p'_{\odd}(w^{2i})
  = p'(w^i).
\]

The second important idea is that the restriction of the degree-$d$
polynomial function $g$ to an affine line can always be written as
\begin{align*}
  g(a_1 + t b_1, \dots, a_q + t b_q) = \sum_{j \in [0,d]} t^j
  g_j(a_1,b_1,\dots,a_q,b_q)
\end{align*}
for some fixed component functions $g_j$ which also have degree at most $d$.
This implies that the even and odd parts of
\begin{align*}
\sumPoly
  &= g(f_{1,\even}(x^2)+xf_{1,\odd}(x^2), \dots,
  f_{q,\even}(x^2)+xf_{q,\odd}(x^2)) \\
  &= \sum_{j \in [0,d]} x^j g_j(f_{1,\even}(x^2), f_{1,\odd}(x^2),
  \dots, f_{q,\even}(x^2), f_{q,\odd}(x^2))
\end{align*}
are
\begin{align*}
  \sumPoly_{\even} &= \sum_{j \in \left[0,\floor{\frac{d}{2}}\right]}
  x^{j} g_{2j}(f_{1,\even}(x), f_{1,\odd}(x), \dots, f_{q,\even}(x),
  f_{q,\odd}(x)), \\
\sumPoly_{\odd} &= \sum_{j \in \left[0,\floor{\frac{d}{2}}\right]}
  x^{j} g_{2j+1}(f_{1,\even}(x), f_{1,\odd}(x), \dots,
  f_{q,\even}(x), f_{q,\odd}(x)).
\end{align*}

We next describe the DGM protocol, however, without reference to any
concrete polynomial commitment scheme and in a way that makes the black-box applicability of Gemini explicit.
At a high level, the protocol aims to reduce the claim that $h$
agrees with $P$ on all $w^i$ to the claim that a smaller polynomial
$h'$ agrees with $g(f_{1,\even} + r f_{1,\odd},\allowbreak{}
\dots,\allowbreak{} f_{q,\even}+rf_{q,\odd})$ on all $w^{2i}$.
Here, $r$ is randomness sampled by the verifier.
DGM proceeds in two logical steps.

First, the prover sends oracles to the low-degree extension polynomials
\begin{equation}
  h'_j = g_j(f_{1,\even},f_{1,\odd},\dots,f_{q,\even},f_{q,\odd})
  \rem (x^{2^{m-1}}-1)
  \label{eq:decompForm}
\end{equation}
for $j \in [0,\allowbreak{}d]$.
The verifier is supposed to check these oracles for consistency with
$h$ by testing, at a random point,
\begin{equation}
  \sum_{j \in [0,d]} x^j h'_{j}(x^2) =^? h(x).
  \label{eq:dgmFlaw}
\end{equation}
This is supposed to establish that the polynomials $\sum_{j \in
\left[0,\floor{\frac{d}{2}}\right]} x^j h'_{2j}(x)$ and $\sum_{j \in
\left[0,\floor{\frac{d}{2}}\right]} x^j h'_{2j+1}(x)$ are the even
and, respectively, odd part of $h$.
Note that this check \emph{does not work}---it is neither complete nor sound.
\emph{\cref{eq:dgmFlaw} does not actually hold as a polynomial
identity because its left and right side have different degrees}. 

Then, the protocol goes on to convince the verifier that these
supposed even and odd parts of $h$ agree with $\sumPoly_{\even}$ and
$\sumPoly_{\odd}$ on all $w^{2i}$.
Since they were constructed as $\sum_{j \in
\left[0,\floor{\frac{d}{2}}\right]} x^j h'_{2j}$ and $\sum_{j \in
\left[0,\floor{\frac{d}{2}}\right]} x^j h'_{2j+1}$, this will
necessarily be the case if the $h'_j$ are well-formed, i.e., were
indeed computed as per \cref{eq:decompForm}.
The verifier now samples $r$, both prover and verifier set $h' =
\sum_{j \in [0,d]} r^j h'_j$ and the protocol is run recursively.
If the recursive protocol succeeds, the verifier knows that the polynomial
\begin{align}
  g(f_{1,\even} + r f_{1,\odd}, \dots, f_{q,\even} + rf_{q,\odd})
  = \sum_{j \in [0,d]} r^j g_j(f_{1,\even}, f_{1,\odd}, \dots,
  f_{q,\even}, f_{q,\odd}) \label{eq:g_j}
\end{align}
agrees with $h'$ on all $w^{2i}$.
By Schwartz-Zippel, this shows that each $h'_j$ agrees with its
respective $g_j$ term on all $w^{2i}$ except with negligible
probability over the choice of $r$.
Assuming that the $h'_j$ are also of degree at most $2^{m-1}-1$, this
ensures that they are correctly built.

After $m$ recursive steps, the remaining claim only includes constant
polynomials.
Because each $f_i$'s even and odd parts were folded $m$ times---say,
with $r_1,\allowbreak{}\dots,\allowbreak{}r_m$---, the resulting
constant is exactly $\mlin[f_i](r_1,\allowbreak{}\dots,\allowbreak{}r_m)$.
That is, the constants are evaluations of the multilinear polynomials
with the same coefficient vectors as the $f_i$.
At that point, Gemini can be used as a black box.

\subsection{Corrected Protocol}\label{sec:dgm_sumcheck}

If we define
\begin{align}
  h_{2j} &= (x^j h'_{2j}) \rem (x^{2^{m-1}}-1) \nonumber\\
  h_{2j+1} &= (x^j h'_{2j+1}) \rem (x^{2^{m-1}}-1),
  \label{eq:h_j}
\end{align}
then the degree-corrected form of \cref{eq:dgmFlaw} is
\begin{equation}
  \sum_{j \in \left[0,\floor{\frac{d}{2}}\right]} h_{2j}(x^2) + x
  h_{2j+1}(x^2) = h(x).
  \label{eq:dgmFix}
\end{equation}
Both sides clearly have degree-$(2^m-1)$ polynomials that agree on all $w^i$.
Thus, to fix the protocol, we propose the prover interpolate each
$h_j$ from its values on the $w^{2i}$ and send it along with $h'_j$.
The verifier may check $h_j$ against $h$ via \cref{eq:dgmFix} and
then, as before, sample randomness and use $h'_j$ to construct the
recursive claim.

For our approach to work, the verifier must also confirm that
$\deg(h'_j) \leq 2^{m-1}-1$ and that $h_j$ and $h'_j$ are consistent
with each other, i.e., that each $h_j$ is indeed constructed as per
\cref{eq:h_j}.
In fact, it is always valid to replace $x^j$ in \cref{eq:h_j} with
$x^{j \rem 2^{m-1}}$ since the remainder modulo $x^{2^{m-1}}-1$ is the same.
Then, it suffices for the prover to show that $\deg(h_j) \leq
2^{m-1}-1$ and that there exist $q_{2j},\allowbreak{} q_{2j+1}$ with
degree at most $j \rem 2^{m-1} - 1$ such that
\begin{align}
  x^{j \rem 2^{m-1}} h'_{2j} &= h_{2j} + q_{2j} \cdot (x^{2^{m-1}}-1)
  \nonumber\\
  x^{j \rem 2^{m-1}} h'_{2j+1} &= h_{2j+1} + q_{2j+1} \cdot (x^{2^{m-1}}-1).
  \label{eq:q_j}
\end{align}
Since on the left-hand side of \cref{eq:q_j} the lowest $j \rem
2^{m-1}$ coefficients are all 0, $q_{2j}$ and $q_{2j+1}$ must contain
just the lowest $j \rem 2^{m-1}$ coefficients of $h_{2j}$ and,
respectively, $h_{2j+1}$. When assuming that $\deg(h_{2j}),\allowbreak{} \deg(h_{2j+1}) \leq
2^{m-1}-1$, we observe that the same coefficients must be the top
coefficients of $h'_{2j}$ and $h'_{2j+1}$.
The prover will send these coefficients in the clear and give an
explicit degree check on every $h_j$.
This allows testing \cref{eq:q_j} at a random point, thereby
simultaneously confirming that $h_j$ and $h'_j$ are consistent in the
sense of \cref{eq:h_j} and that $\deg(h'_j) \leq 2^{m-1}-1$.

For a degree check, we use the standard fact that any polynomial $p$
has degree at most $2^{m-1}-1$ if and only if the rational function
$x^{2^{m-1}-1} p(x^{-1})$ is a polynomial on $\FF \setminus \{0\}$.

The full protocol goes as follows.

\begin{protocol}{Univariate domain identity (reduction step)}{dgm}
  \textbf{Claim:} $g(f_1,\allowbreak{} \dots,\allowbreak{} f_q) \rem
  (x^{2^m}-1) = h.$
  \begin{enumerate}
    \item The prover sends, for $j \in [0,\allowbreak{}d]$,
      \begin{align*}
        h'_j &= g_j(f_{1,\even}, f_{1,\odd}, \dots, f_{q,\even},
        f_{q,\odd}) \rem (x^{2^{m-1}}-1) \\
        h_j &= (x^{\floor{\frac{j}{2}}} g_j(f_{1,\even}, f_{1,\odd},
        \dots, f_{q,\even}, f_{q,\odd})) \rem (x^{2^{m-1}}-1) \\
        c_j &= x^{2^{m-1}-1} h_j(x^{-1}) \\
        q_j &= \sum_{\ell \in \left[0,\floor{\frac{j}{2}} \rem
        2^{m-1} -1\right]} h_j[\ell] x^\ell \quad\text{(sent as
        $\textstyle\floor{\frac{j}{2}} \rem 2^{m-1}$ coefficients)}.
      \end{align*}
    \item The verifier checks, by querying at a random non-zero point,
      \begin{align*}
        c_j(x) &= x^{2^{m-1}-1} h_j(x^{-1}) \,\forall j \in [0,d] \\
        h(x) &= \sum_{j \in \left[0,\floor{\frac{d}{2}}\right]}
        h_{2j}(x^2) + x h_{2j+1}(x^2) \\
        x^{\floor{\frac{j}{2}} \rem 2^{m-1}} h'_j(x) &= h_j(x) +
        q_j(x) \cdot (x^{2^{m-1}}-1) \,\forall j \in [0,d].
      \end{align*}
    \item The verifier sends $r$, chosen at random.
    \item Both prover and verifier define $h' = \sum_{j \in [0,d]} r^j h'_j.$
  \end{enumerate}
  \textbf{New claim:}
  \[
    g(f_{1,\even} + r f_{1,\odd},\allowbreak{} \dots,\allowbreak{}
    f_{q,\even} + r f_{q,\odd}) \rem (x^{2^{m-1}}-1) = h'.
  \]
\end{protocol}

Note that $h'$ does not have to be sent as its own oracle because the
next round's query to $h'$ can be implemented via queries to the
components $h'_j$.

\begin{proposition}\label{prop:dgm}
  \cref{proto:dgm} is a reduction from the relation
  \[
    R = \left\{
      \left(\oracle{f_1}, \dots, \oracle{f_q}, \oracle{h}; f_1,
      \dots, f_q, h\right)
      \,\middle|\,
      g(f_1,\dots,f_q) \rem (x^{2^m}-1) = h
    \right\}
  \]
  to (after $m$ steps)
  \begin{align*}
    R' = \BigSetInAlign{
      \left(\oracle{f_1}, \dots, \oracle{f_q}, r_1, \dots, r_m, s;
      f_1, \dots, f_q\right)
    }{
      g(\mlin[f_1](r_1, \dots, r_m), \dots, \mlin[f_q](r_1, \dots, r_m)) = s
    },
  \end{align*}
  where $s$ is a constant, with
  \begin{itemize}
    \item perfect completeness,
    \item soundness error in $O(\frac{dm\degBound}{|\FF|})$ (where
      $\degBound$ is a global upper bound on any oracle polynomial's degree),
    \item prover running time in $O(\binom{d+2q}{2q} d 2^m)$,
    \item verifier running time in $O(d^2 m)$,
    \item $m$ rounds,
\item $O(d^2 m)$ field elements sent by the prover,
\item $O(dm)$ oracles sent by the prover,
\item $O(dm)$ oracle queries made by the verifier.
  \end{itemize}
\end{proposition}

\begin{proof}
  The properties besides completeness, soundness and prover run time
  are obvious from the description of \cref{proto:dgm}.

  \paragraph{Completeness.}
  Completeness of the $m$-step reduction follows from the
  completeness of each step.
  Suppose $g(f_1,\allowbreak{} \dots,\allowbreak{} f_q) \rem
  (x^{2^m}-1) = h$ and the prover is honest.
  Then the first verifier check passes because it only tests the
  definition of each $c_j$.
The second and third check pass since they test the polynomial
  identities from \cref{eq:dgmFix,eq:q_j}.
  Finally, the new claim must be correct due to \cref{eq:g_j} and the
  fact that each $h_j$ agrees with $g_j(f_{1,\even},\allowbreak{}
    f_{1,\odd},\allowbreak{} \dots,\allowbreak{}
  f_{q,\even},\allowbreak{} f_{q,\odd})$ on all $w^{2i}$ by definition.

  \paragraph{Soundness.}
  Suppose $g(f_1,\allowbreak{} \dots,\allowbreak{} f_q) \rem
  (x^{2^m}-1) \neq h$ and the verifier accepts.
  We prove that the soundness error is at most
  $\frac{m(d+(2d+3)\degBound)}{|\FF|-1}$ by induction over $m$.
  For $m=1$, the reduction takes just one round to reduce the claimed
  polynomial identity on $x=\pm 1$ to identity on $x=1$.
In this case, no coefficients are sent for any $q_j$, so $q_j=0$.
  Since the verifier accepts, the checks in step 2 must succeed.
  Let $E$ be the event that all of the $2d+3$ checked equations
  indeed hold as polynomial identities.
  Note that $\Pr[E]$ is 1 minus the probability that any does not
  hold as an identity but holds at the independently sampled point.
  For each equation individually, that probability is at most
  $\frac{\degBound}{|\FF|-1}$ since any two univariate polynomials
  with degree at most $\degBound$ agree on at most $\degBound$ points.
  By a union bound, $\Pr[E] \geq 1 - \frac{(2d+3)\degBound}{|\FF|-1}$.
  Then, firstly, $h_j=h'_j$ and, secondly, $\sum_{j \in
  \left[0,\floor{\frac{d}{2}}\right]} h_{2j}$ and $\sum_{j \in
  \left[0,\floor{\frac{d}{2}}\right]} h_{2j+1}$ are the even and,
  respectively, odd part of $h$.
  Since the starting claim is false, the even and odd part of $h$ do
  not agree with the even and odd part of
  $g(f_1,\allowbreak{}\dots,\allowbreak{}f_q)$ on $x=1$.
  Therefore, it certainly cannot be that for all $j$, $h'_j =
  g_j(f_{1,\even},\allowbreak{} f_{1,\odd},\allowbreak{}
  \dots,\allowbreak{} f_{q,\even},\allowbreak{} f_{q,\odd})$.
  This implies that the recursive claim is false except for the
  probability that two degree-$d$ univariate polynomials agree at the
  independently sampled point $r$.
  That probability is at most $\frac{d}{|\FF|}$. Together, the soundness error is at most $\Pr[E] \frac{d}{|\FF|} +
  1 - \Pr[E] \leq \frac{d+(2d+3)\degBound}{|\FF|-1}$.

  For $m \geq 2$, since the verifier's checks in step 2 succeed, the
  checked equations hold as polynomial identities with probability
  $\Pr[E] \geq 1 - \frac{(2d+3)\degBound}{|\FF|-1}$.
  Then, firstly, each $h_j$ agrees with $x^{\floor{\frac{j}{2}}}
  h'_j$ on the halved domain and, secondly, $\sum_{j \in
  \left[0,\floor{\frac{d}{2}}\right]} h_{2j}$ and $\sum_{j \in
  \left[0,\floor{\frac{d}{2}}\right]} h_{2j+1}$ are the even and,
  respectively, odd part of $h$.
  Since the starting claim is false, they do not agree with the even
  and odd part of $g(f_1,\allowbreak{}\dots,\allowbreak{}f_q)$ on the
  halved domain.
  Therefore, it certainly cannot be that for all $j$, $h'_j$ agrees
  with $g_j(f_{1,\even},\allowbreak{} f_{1,\odd},\allowbreak{}
  \dots,\allowbreak{} f_{q,\even},\allowbreak{} f_{q,\odd})$ over the
  halved domain.
  This implies that the recursive claim is false except with
  probability at most $\frac{d}{|\FF|}$ over the choice of $r$.
  Let $E'$ be the event that the recursive protocol succeeds despite
  a false claim.
  By the induction hypothesis, $\Pr[E'] \leq (m-1)
  \frac{d+(2d+3)\degBound}{|\FF|-1}$.
  All in all, the soundness error is at most
  \begin{align*}
    &\Pr[E] (\Pr[\text{new claim true}] + \Pr[\text{new claim false}]
    \Pr[E']) \\
    &+ 1 - \Pr[E] \\
    \leq& \frac{d}{|\FF|} + \Pr[E'] + \frac{(2d+3)\degBound}{|\FF|-1} \\
    \leq& m \frac{d+(2d+3)\degBound}{|\FF|-1}.
  \end{align*}

  \paragraph{Prover time.}
  First, it takes time in $O(q2^{m-1})$ for the prover to compute all
  values $f_{k,\even}(w^{2i})$ and $f_{k,\odd}(w^{2i})$ for
  $k\in[q],\allowbreak{}i\in[0,\allowbreak{}2^{m-1}-1]$.
  Then, the prover needs to compute $2^{m-1}$ evaluations the $d+1$
  $2q$-variate degree-$d$ polynomials $g_0,\allowbreak{}
  \dots,\allowbreak{} g_d$.
  With the bound $\binom{d+2q}{2q}$ for the time to compute each
  evaluation, this takes time in $O(\binom{d+2q}{2q} d 2^{m-1})$.
  At most $2^{m-1}$ extra steps are needed to compute all domain
  values of $h_j$ and $c_j$.
  To compute the coefficients of $q_j$, the following well-known
  ``DFT'' formula can be used: \begin{align*}
    \sum_{i \in [0,2^{m-1}-1]} w^{-2i\ell} h_j(w^{2i})
    &= \sum_{i \in [0,2^{m-1}-1]} w^{-2i\ell} \sum_{k \in
    [0,2^{m-1}-1]} h_j[k] w^{2ik} \\
    &= \sum_{k \in [0,2^{m-1}-1]} h_j[k] \sum_{i \in [0,2^{m-1}-1]}
    w^{2i(k-\ell)} \\
    &= 2^{m-1} h_j[\ell]
  \end{align*}
  where the last equality used that $\sum_{i \in [0,2^{m-1}-1]}
  (w^{2(k-\ell)})^i$ is 0 except when $(k-\ell) \rem 2^{m-1}=0$,
  which is the case only if $k=\ell$.
  That is, reading from right to left, the coefficient $h_j[\ell]$
  can be computed by summing $2^{m-1}$ scaled evaluations of $h_j$
  and scaling the result.
  Since there are up to $O(d^2)$ coefficients, this gives a running
  time in $O(d^2 2^{m-1})$.
  So far, we have prover time in $O((q + \binom{d+2q}{2q} d + d^2)
  2^{m-1}) = O(\binom{d+2q}{2q} d 2^{m-1})$ for the first round.
  After preparing the values of all $f_{i,\even} + r f_{i,\odd}$ in
  another $2^{m-1}$ steps, the subsequent round will take the same
  computations as the first round, except for the halved domain size.
  Summing over all rounds, we obtain running time in
  $O(\binom{d+2q}{2q} d 2^{m})$ overall.

  \paragraph{}
  This completes the proof of \cref{prop:dgm}.
\end{proof}

For a complete PIOP for domain identity, prover and verifier would
first run \cref{proto:dgm} $m$ times to obtain a claim of the form
$g(\mlin[f_1](r_1,\allowbreak{} \dots,\allowbreak{}
  r_m),\allowbreak{} \dots,\allowbreak{}
\mlin[f_q](r_1,\allowbreak{}\dots,\allowbreak{}r_m)) = s$.
They would then construct a batched $\mlin$ evaluation instance by
the prover sending claimed evaluations of the $\mlin[f_i]$, the
verifier sending a randomly sampled point $t$ and then setting $f^* =
\sum_i t^i \mlin[f_i]$.
The final claim involving $f^*$ would be handled by running Gemini.

\paragraph{Sumcheck.}
Similarly to \cite{aurora}, sumcheck can be obtained from a domain
identity check by querying the authenticated remainder polynomial at 0.
For example, the prover could provide a decomposition of $h$ as $h(0)
+ x h'(x)$.
The resulting sumcheck's costs are shown in the fourth row of
\cref{tab:summary}.
By comparison, this method is clearly less efficient than using the
multivariate sumcheck protocol in combination with the adaptor from
\cref{sec:mlex2unex} due to \cref{proto:dgm}'s extra oracles and field elements.

\paragraph{Round reduction.}
Similarly to \cref{sec:round_reduction}, the domain identity protocol
can be early-stopped and composed with a separate single-round domain
identity check like Aurora's \cite{aurora}.
We summarize the resulting round-reduced sumcheck protocol's
performance in the fifth row of \cref{tab:summary}.
The remarks from \cref{sec:round_reduction} apply.

\printbibliography[heading=bibintoc]

\end{document}